\documentclass[11pt]{article}

\usepackage[margin=1in]{geometry}
\usepackage{lmodern}
\usepackage{url}
\usepackage{amsmath,amssymb}
\usepackage{float}
\usepackage{tabularx}
\usepackage{booktabs}
\usepackage{graphicx}
\usepackage{makecell}
\usepackage[hidelinks]{hyperref}
\usepackage{apacite}

\title{Evaluating LLMs for Translating Legacy Fortran Atmospheric Physics to GPU-Accelerated JAX}
\author{
  Iris Linck\thanks{Corresponding author: ilinck@ucar.edu},
  John Dennis, John Schreck,\\
  Karen Stengel and David John Gagne II\\[0.5em]
  {\small NSF National Center for Atmospheric Research}
}
\date{}
\usepackage{microtype}
\renewenvironment{abstract}
  {\section*{Abstract}\normalfont\normalsize}
  {\par}
\begin{document}
\maketitle

\begin{abstract}
Legacy Fortran physics packages in production atmospheric models such as CESM, WRF, and E3SM remain a major barrier to GPU modernization, while manual porting requires substantial accelerator-programming expertise. We present an LLM-Assisted Fortran Translation (LAFT), an agentic framework for converting Fortran atmospheric physics parameterizations into GPU-accelerated JAX while preserving numerical correctness and exposing optimization opportunities to XLA. LAFT coordinates source-code analysis, JAX translation, error repair, runtime validation, and performance profiling. The framework combines a deterministic front end that extracts Fortran metadata and generates bridge code for column-major to row-major layout conversion, structured prompt policies that guide LLM code generation, validation against the Fortran reference, and profiling-guided feedback that identifies inefficient execution patterns in the generated kernels. We evaluate LAFT in an initial case study using the Kessler warm-rain microphysics scheme and four LLMs: Claude Sonnet 4.6, GPT-5.4 Thinking, Gemini 3.1 Pro, and Qwen2.5-32B. The validated translations reproduce the Fortran reference outputs to near floating-point precision, while performance differences arise primarily from the execution designs generated by each LLM. The fastest translated JAX implementation, generated using Gemini, achieves up to a 357× speedup over serial Fortran in cached GPU execution at $10^{6}$ atmospheric columns. The results show that LLM-assisted translation can produce highly accelerated JAX implementations that remain consistent with the Fortran reference, but only when embedded in a workflow that combines deterministic interface generation, automated validation, repair, and hardware-level profiling. By coupling LLM translation with reproducible validation and profiling, LAFT provides a practical path toward modernizing legacy atmospheric physics codes for GPU-accelerated Earth-system modeling.
\end{abstract}

\section*{Plain Language Summary}
Weather and climate models rely on many older computer programs written in Fortran and designed for traditional CPUs. Modern supercomputers increasingly use GPUs, which can be much faster but often require substantial code changes. Manually rewriting scientific codes for GPUs is difficult, time-consuming, and requires specialized programming expertise. This study presents LAFT, a framework that uses large language models and automated agents to help translate Fortran atmospheric physics code into JAX, a Python-based system that can run efficiently on GPUs. LAFT uses agents to generate translations, fix errors, validate results against the original Fortran code, and profile GPU performance. A key part of the framework is a bridge layer that handles differences in how Fortran and Python organize array data in memory, allowing the translated code to focus on the scientific calculations. We demonstrate the framework using the Kessler microphysics scheme, a simplified model of warm-rain processes. The study shows how LLM-assisted translation, combined with automated repair, validation, and performance profiling, can support modernization of legacy atmospheric model components for GPU-based supercomputers.

\section{Introduction}
\label{sec:introduction}
Numerical weather prediction and climate modeling depend on large scientific software systems that have evolved over several decades. Many widely used Earth-system models, including CESM, WRF, and E3SM, rely on atmospheric physics packages written in Fortran. These codes encode mature scientific knowledge and have been extensively tested in production modeling environments, but many were designed originally for CPU-based execution. As high-performance computing systems increasingly rely on GPU-accelerated architectures, legacy Fortran physics packages have become a major barrier to performance portability and scientific software modernization. Manually porting such codes to GPUs requires expertise in accelerator programming, memory-layout management, compiler behavior, and performance tuning, creating a bottleneck for modeling groups that must modernize code while preserving numerical correctness. 
Modern array-based programming systems such as JAX offer a promising path toward GPU acceleration for scientific kernels. JAX provides just-in-time compilation and accelerator execution through XLA~\cite{noauthor_jax-mljax_2026}, making it an attractive target for translating loop-based Fortran physics code into GPU-executable array programs. However, translating Fortran scientific code to JAX is not merely a syntactic conversion. The two programming models differ in memory layout, indexing conventions, mutation semantics, control-flow structure, numerical behavior, and performance expectations. A direct translation that mirrors Fortran syntax may preserve the apparent computation but perform poorly on GPUs, while an aggressive rewrite may expose more parallelism at the risk of changing the scientific behavior. A useful modernization workflow must therefore address correctness and performance together. Although JAX also supports automatic differentiation, this paper focuses on numerical correctness and GPU performance rather than differentiability.

Automated code translation has long been studied in software engineering, from rule-based transformation systems such as TXL~\cite{cordy_txl_1988} to systems based on static analysis, intermediate representations, and domain-specific rewrite rules~\cite{khalafinejad_rule-based_2010,emani_dbridge_2017,noor_translation_2020}. These approaches provide explicit control over translation behavior, but they require substantial manual engineering and often depend on source- and target-specific transformation logic. Large language models offer a different opportunity because they can use broader context and often generate more idiomatic target-language code. Recent studies have explored LLM-assisted code translation and shown that prompt-provided context can improve translation outcomes~\cite{pan_stelocoder_2023,pan_lost_2024}. However, LLM-generated translations are not guaranteed to compile, preserve semantics, or perform efficiently. Empirical studies report compilation errors, runtime errors, functional errors, missing dependencies, incorrect API mappings, incorrect data types, and other semantic bugs in LLM-based translation~\cite{pan_lost_2024}. These risks are especially important for numerical scientific codes, where small semantic changes can produce scientifically invalid results. Moreover, as this paper shows, even numerically correct translations may execute inefficiently on GPUs if the generated code fails to expose parallelism or produces unfavorable loop and array-update structures. For this reason, recent translation systems increasingly combine LLM generation with program analysis, validation, repair, retrieval, compiler feedback, or agentic refinement~\cite{hong_type-migrating_2024,bhattarai_enhancing_2024,liu_mftcoder_2024,yuan_transagent_2026,dehghan_translating_2025}. A related line of work uses performance profiling as feedback for LLM-driven code generation, feeding hardware metrics back into iterative refinement loops to reduce bottlenecks~\cite{lei_pragma_2025,li_tritonforge_2025}. The common insight is that reliable translation requires more than a single prompt: the model must be embedded in a workflow that provides context, constrains the translation, validates correctness, and uses execution feedback to guide repair and optimization.

In this paper, we introduce LAFT, LLM-Assisted Fortran Translation, an agentic framework for translating Fortran atmospheric physics code into GPU-accelerated JAX. LAFT combines LLM-based code generation with deterministic infrastructure, automated repair, numerical validation, and GPU profiling. The framework is built on the principle that generated scientific code should not be accepted from the LLM directly, but should instead be validated against a reference implementation and refined using execution feedback. It compares each translated JAX implementation against the original Fortran reference and uses GPU- and XLA-level profiling to identify inefficient execution patterns, including loop structure, kernel launches, synchronization, memory movement, and opportunities for XLA fusion. In this workflow, agents are used to generate translations, repair errors, validate numerical behavior, and interpret profiling feedback for subsequent refinement.

This work addresses three research questions:
\begin{description}
    \item[RQ1.] Can current LLMs translate Fortran atmospheric physics kernels into JAX implementations that reproduce reference numerical results?
    \item[RQ2.] Do numerically equivalent LLM-generated translations exhibit comparable GPU performance, or do code design choices such as loop structure, vectorization, and array-update patterns dominate runtime?
    \item[RQ3.] What workflow components, beyond the LLM itself, are necessary to make translation reliable and performance-aware?
\end{description}
To answer these questions, LAFT contributes three evaluations and design elements that distinguish it from general-purpose code-translation frameworks:
\begin{enumerate}
    \item \textbf{Numerical correctness against the Fortran reference.}
    LAFT evaluates each LLM-generated JAX implementation by executing it on the same input state as the original Fortran kernel and comparing the resulting outputs using mean absolute error (MAE). In the Kessler case study, the successful translations reproduce the Fortran reference to near floating-point precision for all evaluated output variables. This result shows that current LLMs can generate JAX implementations that preserve the numerical behavior of the Fortran reference when embedded in a validation-driven workflow.
    \item \textbf{Code design choices determine GPU performance.} The results show that numerically equivalent translations do not necessarily exhibit comparable GPU performance. Although multiple LLM-generated implementations reproduce the Fortran reference, their runtime differs substantially because they expose different computation structures to XLA. In particular, code design choices such as vectorized column execution versus sequential loops, full-array masking versus gather-and-scatter updates, and the use of JAX control-flow primitives determine how much parallelism and fusion XLA can exploit. This shows that, for GPU-oriented scientific translation, correctness is necessary but not sufficient: the generated code must also be structured in a hardware-aware way.
    \item \textbf{Deterministic infrastructure and profiling feedback.} LAFT does not rely on the LLM alone to manage translation complexity. Instead, it uses deterministic components to handle tasks that can be explicitly specified and validated. In particular, LAFT separates memory-layout conversion from scientific computation through a reusable bridge layer. Fortran uses column-major arrays while JAX defaults to row-major. Rather than delegating this conversion to the LLM within each translated kernel, LAFT uses a deterministic front end to extract Fortran metadata and generate a bridge that converts arrays once at the interface. LAFT also includes profiling diagnostics for evaluating the generated JAX implementations. The profiling workflow uses JAX/XLA traces. In GPU-accelerated JAX, performance depends strongly on whether the generated code exposes parallelism and allows XLA to fuse operations efficiently. These diagnostics expose inefficient execution patterns, such as excessive kernel launches, unexpected recompilation, and sequential loop structures that prevent XLA from exploiting parallelism. The resulting profiling information provides  feedback that can be used in subsequent LLM-guided refinements toward more hardware-efficient code.
\end{enumerate}
We evaluate LAFT using the Kessler warm-rain microphysics scheme, a representative atmospheric physics case study. Kessler microphysics is small enough to support detailed validation and profiling, yet rich enough to include features common across atmospheric parameterizations: vertical loops, column-wise structure, conditional updates, and multiple interacting prognostic variables. We compare JAX implementations generated by four LLMs, Claude, GPT, Gemini, and Qwen, against the original Fortran reference using MAE. 

Performance is reported using cached JAX execution times, including
dispatch, device execution, and synchronization, while excluding
compilation and host--device transfers. Bridge overhead is measured
separately through host-to-device transfer, input and output
permutations, and device-to-host transfer, with phase totals reported
for C-contiguous and Fortran-ordered callers. 
This distinction matters because a translated kernel may execute quickly once data is on the GPU, yet practical deployment must also account for layout conversion and transfer costs.

The results show that all four LLM-generated translations closely reproduce the Fortran reference outputs for the evaluated Kessler cases, yet differ substantially in GPU performance. These differences arise from code design choices that affect the parallelism and optimization opportunities available to XLA. Numerical validation establishes agreement with the reference, while hardware-level profiling identifies execution bottlenecks that numerical comparisons cannot reveal. Together, these findings motivate the combination of correctness checks with hardware-aware performance analysis in LLM-assisted scientific-code translation.

Beyond the Kessler case study, this paper contributes a reusable framework for LLM-assisted modernization of legacy atmospheric physics code that combines structured prompt policies, deterministic metadata extraction, bridge-based layout conversion, runtime validation, and hardware-level profiling feedback. LAFT is designed around challenges common to atmospheric physics parameterizations: array-layout mismatch, loop-structured computation, floating-point sensitivity, and GPU execution efficiency.

The remainder of this paper is organized as follows. Section 2 describes the LAFT framework, including the deterministic Fortran front end, bridge generation and validation, the LLM-driven translation workflow, automated validation and repair, and the profiling feedback loop. Section 3 presents the Kessler case study, the LLM-generated implementations evaluated, the hardware and software platform, and the accuracy and performance metrics. Section 4 evaluates the execution design, numerical accuracy, bridge overhead, and cached GPU scaling of the LLM-generated JAX translations. Section 5 compares the best-performing JAX/XLA translation with a manually accelerated OpenACC GPU implementation. Section 6 outlines future work, and Section 7 concludes.

\section{LAFT Framework}
\label{sec:laft-framework}
\begin{figure}[!htbp]
    \centering
    \includegraphics[width=\linewidth]{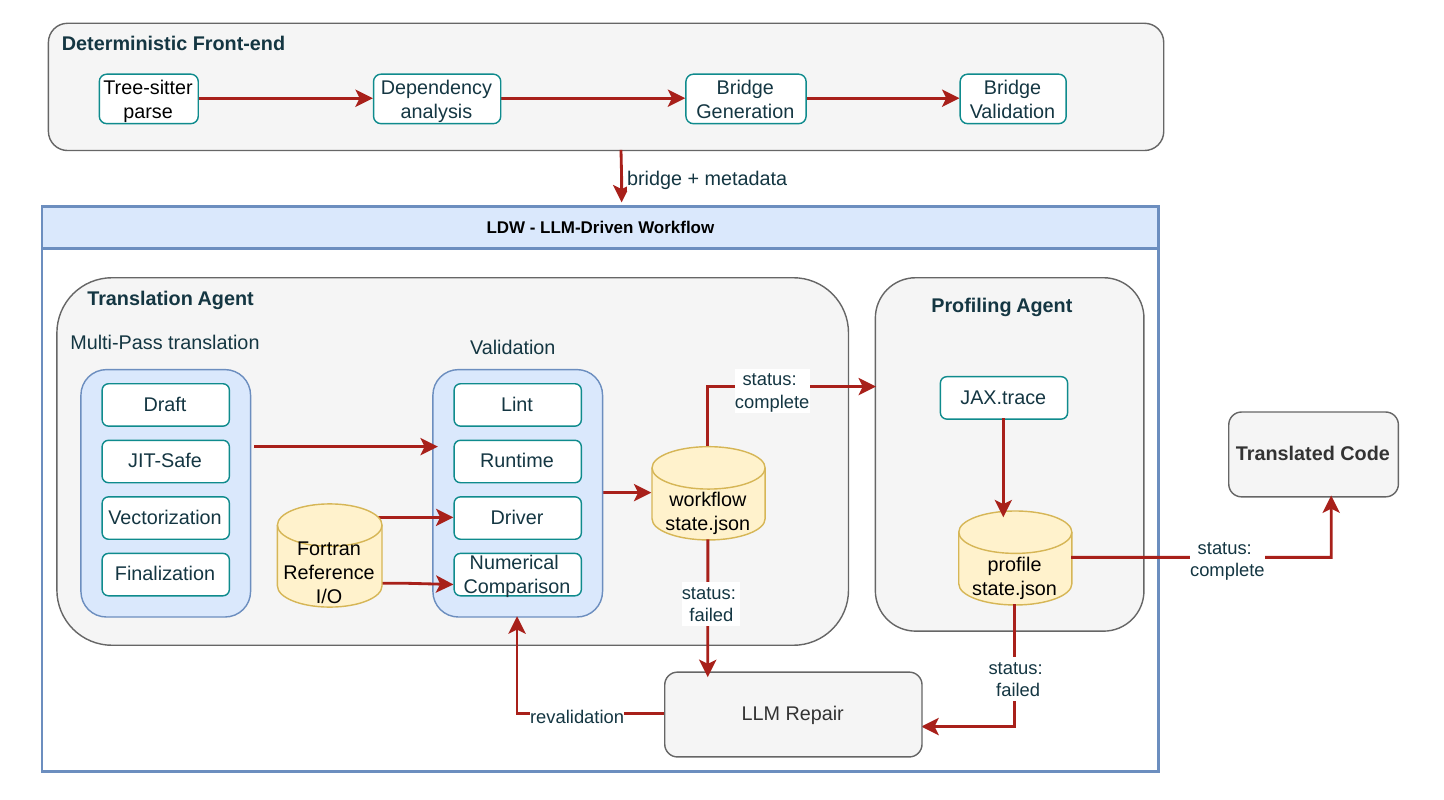}
    \caption{Overview of the LAFT workflow. A deterministic front end extracts Fortran metadata and generates a validated bridge for layout and interface conversion. The bridge and metadata are passed to an LLM-driven workflow, where a translation agent produces JAX code through multiple passes and validates it against the Fortran reference. Validation and profiling outcomes are recorded in workflow state files; failures trigger LLM repair and revalidation, while successful validation and profiling produce the final translated code.}
    \label{fig:laft-pipeline}
\end{figure}

Figure~\ref{fig:laft-pipeline} summarizes LAFT, an LLM-Assisted Fortran Translation framework for converting legacy Fortran atmospheric physics procedures into GPU-accelerated JAX implementations. Rather than treating translation as a single prompt-to-code step, LAFT organizes the process as a staged workflow. 
The framework separates deterministic program analysis and interface generation from LLM-based code generation, then validates and profiles the resulting implementation before accepting it as a candidate translation. This design reflects a central requirement of scientific-code modernization: the translated program must preserve numerical behavior while exposing enough structure for efficient GPU execution. To make the LLM task more reliable, LAFT parses the Fortran source into smaller procedure-level translation units using a Tree-sitter-Fortran parser \cite{stadelman_stadelmanmatree-sitter-fortran_2026} and translates each procedure through a four-pass prompting strategy. This design is motivated by a practical limitation of long-context LLMs: a large context window does not ensure that the model will use all parts of the prompt equally well. Prior work has shown that LLM performance can degrade when relevant information appears in the middle of a long context, even for LLMs designed to accept long inputs~\cite{liu_lost_2024}. This limitation is especially important for scientific-code translation, where critical details such as variable definitions, dependency relationships, array dimensions, mutation behavior, memory-layout assumptions, and control-flow constraints may be buried in a long prompt. If the LLM overlooks any of these details, the generated translation may appear syntactically correct while still producing incorrect numerical behavior or inefficient GPU code. LAFT therefore decomposes translation into focused passes that separate initial code generation, JIT-safety checks, vectorization and XLA-oriented restructuring, and final consistency checks. The goal is not only to reduce prompt length, but also to make each LLM task narrower, more inspectable, and easier to validate.

At a high level, LAFT consists of five stages: (1) deterministic extraction of Fortran procedure metadata, (2) bridge generation for layout and interface conversion, (3) policy-guided LLM translation, (4) correctness validation against the Fortran reference, and (5) hardware-level profiling and feedback. All the stages are described in the following subsections.
\subsection{Deterministic Front End}

\textbf{Tree-sitter Parse:} LAFT front-end converts the input Fortran source into structured metadata before any LLM-based translation occurs. This stage is designed to make the program structure explicit and reproducible, reducing the amount of source-code interpretation delegated to the language model. Rather than relying on line-based pattern matching, LAFT parses the Fortran source using Tree-sitter and a Fortran-specific Tree-sitter grammar ~\cite{stadelman_stadelmanmatree-sitter-fortran_2026} and recovers procedure boundaries, declarations, control-flow constructs, module structure, and source ranges from the resulting abstract syntax tree. Lightweight regular-expression fallbacks are used only for cases where the grammar does not expose a needed detail, such as certain assumed-shape array bounds. The parsing stage isolates each compilation unit, including modules, subroutines, functions, and program blocks, and emits them as standalone source slices with associated metadata. For each procedure, LAFT extracts a typed argument signature that records the declared data type, array rank, symbolic dimension expressions, and intent information where available. At module scope, the front end records use dependencies, module-level variables, accessibility information, and the association between procedures and their parent modules. It also preserves OpenMP directives, since these annotations can signal loop-level parallelism and may inform later vectorization decisions.

\textbf{Dependency Analysis:} After parsing, LAFT analyzes dependencies at both the module and procedure levels. At the module level, it builds a dependency graph from the parser metadata, where each node is a Fortran module and each edge represents an internal \texttt{USE} dependency within the translation corpus. The graph is used to identify circular dependencies and to assign dependency levels, so modules with no internal prerequisites can be translated first and dependent modules can be translated only after their required modules are available.

\textbf{Bridge Generation:} Unlike the compute kernels, the bridge is not produced by the language model. It is generated deterministically from the structured metadata emitted
by the front end. Its purpose is to isolate data-layout conversion from scientific computation, so that the LLM-generated JAX kernels can assume a single array
convention and do not need to implement Fortran/Python layout handling themselves. Callers pass ordinary NumPy arrays in the Fortran-facing index order, such as \texttt{(ncol, nz)}, and the bridge transfers them to the device in that same index order, with no host-side transposition. In the experiments reported here, the caller is a Python driver whose arrays use NumPy's default row-major memory order, so they are transferred as they are. Arrays stored in column-major memory order, as a Fortran host model would provide, currently incur an additional host-side copy before transfer (Table~\ref{tab:bridge_overhead}). For each procedure with array arguments, the generator emits a private JAX-jitted device
wrapper around the translated {procedure}\texttt{\_core} function. Inside this wrapper, every rank-two or higher array has its axes reversed with jnp.transpose before the core
call, converting the Fortran-facing array order into the layout expected by the JAX kernel. After the core returns, output arrays are transposed back inside the same jitted region before being returned to the host. Because the transposes are inside the JIT boundary, XLA can optimize the layout permutation together with the kernel computation and fuse it with adjacent device operations. One-dimensional arrays pass through unchanged, scalars bypass array conversion, and integer or logical scalar arguments are marked static when required by
JAX tracing. 
Real-valued arrays are converted to 64-bit floating point to match
the Fortran double-precision reference. The current generator applies
the same conversion to all non-character arrays and does not yet
preserve integer or logical array types, which the Kessler interface
does not use. Integer and logical scalars retain their types. 
The bridge’s public contract remains Fortran-facing: callers pass and receive standard NumPy arrays in the original Fortran index order. Device-resident outputs are fetched back to the host with a single batched jax.device\texttt{\_get}, which returns all numeric outputs through one synchronization point. Character arguments remain host-side and pass through the bridge unchanged, since strings cannot cross the JAX JIT boundary. Module variables are represented explicitly as INOUT arguments: they appear in the bridge signature, are passed to the device wrapper when numeric, and are returned among the bridge outputs so that the driver manages module state explicitly rather than relying on hidden global state.

\textbf{Bridge Validation:} The bridge layer is validated independently from the scientific Fortran comparison. These tests verify the correctness of the layout and interface contract implemented by the bridge, including round-trip layout fidelity, correct transposition between Fortran and JAX layouts, preservation of expected shapes and memory contiguity, propagation of error/status fields, and handling of module-state variables. The tests use synthetic, spatially uniform sample fields and compare the output of the bridged call against a direct invocation of the generated JAX kernel. Thus, this validation checks whether the bridge introduces any layout, interface, or numerical discrepancy relative to the JAX kernel itself; it does not establish scientific equivalence with the original Fortran implementation. Scientific equivalence is assessed separately in the end-to-end validation stage.

\subsection{LLM-Driven Workflow}
LAFT implements the LLM-driven workflow (LDW) as a sequential, artifact-mediated multi-agent workflow. The Translation Agent (TA) first generates and repairs a candidate JAX implementation until it passes the validation ladder. The validated translation is then handed off to the Profiling Agent (PA), which profiles GPU/XLA behavior and may apply performance-oriented repairs. Because such repairs can affect correctness, any profiling-driven modification is routed back through the same validation ladder before the code is accepted as final. The agents do not coordinate through direct dialogue; instead, they coordinate through durable artifacts, including validated code snapshots, validation reports, profiling reports, and workflow-state files. Both agents are implemented using command-line coding assistants, which are tools that wrap an LLM with the ability to read and write repository files, run shell commands, and submit or monitor batch jobs. Each run of the LDW is driven by a single orchestrating assistant that performs translation, validation, and repair. We consider three such self-contained agentic configurations: Claude Code backed by Claude, OpenAI Codex backed by GPT, and Gemini CLI backed by Gemini. Each assistant is given the same model-agnostic workflow specification, allowing the workflow to remain fixed while the underlying model varies. The orchestrator normally generates the translation itself; the one exception is the Qwen2.5-32B configuration, in which the orchestrator obtains the draft translation from the Qwen API but still runs validation and repair. Because only these three assistants are available as agentic command-line tools in our environment, Qwen2.5-32B participates solely as a translation engine and never as an orchestrator.
The translation and profiling stages may be executed by the same coding assistant or by different assistants. This separation allows LAFT to evaluate translation quality independently from profiling and feedback generation, while still supporting an integrated end-to-end workflow. We describe each LDW component below.

\subsubsection{Multi-pass LLM Translation.}
For each Fortran procedure, the TA assistant consumes the prompt chain generated by the deterministic front end and produces the JAX/Python translation through a sequence of refinement passes. Array-based Fortran procedures are translated using four passes: an initial draft, a JIT-safety pass, a vectorization and control-flow pass, and a finalization pass that enforces data-type consistency, performs a self-check, and adds documentation. Each pass receives the output of the previous pass, making the refinement cumulative. Scalar-only procedures, such as \texttt{kessler\_init}, are translated in a single pass into plain Python, avoiding unnecessary JAX or accelerator overhead. After writing the translated code, the assistant runs a sequence of automated validation gates: static linting, runtime validation, end-to-end driver execution through the bridge, and numerical comparison against the original Fortran reference. Each validation step is executed as a batch job on the target system. The assistant records the job identifier, waits for completion, checks the logs, and verifies that the result files are fresh and traceback-free before proceeding. Agreement with the Fortran reference is established only in the final numerical comparison stage; earlier gates check structural, import, runtime, and driver-level correctness. 
One model in our study, Qwen2.5-32B, was not available as an agentic command-line assistant in our environment and therefore did not orchestrate its own translation workflow. Instead, Claude Code executed the workflow on its behalf: it issued the four-pass translation prompts to Qwen2.5-32B through its API, wrote the returned code to disk, and then ran the same validation sequence used for the other models. Claude Code also applied any repairs required to pass the validation gates. Because the repair step was not authored by Qwen2.5-32B itself, we report the Qwen2.5-32B results as reflecting a Qwen-generated translation code subsequently repaired through Claude Code.

\subsubsection{Validation Workflow}
\label{sec:validation}
After translation, LAFT validates the generated implementation before it is used for performance analysis. The validation workflow includes static linting, runtime execution checks, driver execution, and numerical comparison against the Fortran reference. For each validation step, the TA launches a PBS job on the target HPC system, records the submitted job identifier, waits for completion, and checks the resulting logs and output files before advancing to the next gate. Together, the validation stages form an increasing-fidelity ladder. Static linting checks structural compliance with the LAFT translation policy; runtime validation checks importability, execution, JIT compatibility, and finite outputs on synthetic inputs; the end-to-end driver runs the translated workflow on real shared inputs; and the numerical comparison stage measures agreement with the Fortran reference. Only the final stage establishes scientific correctness with respect to the reference implementation. Each validation stage is described below.

\textbf{Linting Stage:} The linting stage statically checks the generated code against the LAFT bridge-mode translation policy. For array-based procedures, the linter requires the use of JAX constructs, including \texttt{jax} and \texttt{jax.numpy}, and rejects NumPy usage inside the translated implementation. It also enforces the expected two-layer organization consisting of a wrapper and a JIT-compiled \texttt{\_core} function. Scalar-only procedures are treated differently: they are expected to remain plain Python routines, without JAX imports or a \texttt{\_core} layer, to avoid unnecessary accelerator overhead. The linter also enforces the bridge layout policy. Since layout conversion is handled by the generated bridge, translated kernels must operate under the row-major layout assumption and must not perform additional in-module layout conversions such as \texttt{swapaxes}. The linter further checks that I/O does not appear inside the JIT-compiled \texttt{\_core} function, and that wrapper-level I/O is allowed only for procedures explicitly classified as I/O-bearing. It also prevents strings from crossing the JIT boundary into \texttt{\_core}, since such values are not suitable for JAX compilation. Finally, the linter verifies that the wrapper signature matches the expected LAFT interface. This interface is derived from the original Fortran arguments, excluding scalar output-only flags where appropriate, and including the module variables used by the procedure. In this way, linting catches structural and policy violations before runtime validation or numerical comparison are attempted.

\textbf{Runtime validation.} The runtime-validation stage performs a smoke test and JIT-compatibility check on the translated module using synthetic inputs. It first imports the generated module and then executes two validation paths. The first path calls the bridge entry point, using dummy arrays in the Fortran layout, \texttt{(ncol, nz)}, thereby exercising the full stack, including bridge-level layout conversion. The second path calls the generated core routine directly with the same inputs transposed into the JAX layout, \texttt{(nz, ncol)}, and compiles the call using \texttt{jax.jit}. The validator records whether the module imports successfully, whether the wrapper and core calls execute without exception, and whether all returned numerical values are finite. Results are written per procedure to a runtime-validation report. Runtime validation also reinforces the bridge-mode policy. If the expected generated bridge is missing, the validator fails. Scalar arguments are passed as static arguments when required by \texttt{jax.jit}, and string arguments are rejected before crossing into the JIT-compiled core. The core/JIT test is skipped for scalar-only procedures that do not define a \texttt{\_core} function and for wrapper-only I/O procedures. A procedure passes runtime validation only when the required execution paths complete successfully.

\textbf{End-to-end JAX driver:} After runtime validation, LAFT executes the translated code through an end-to-end JAX driver that mirrors the structure of the Fortran reference driver. The driver reads the atmospheric input state from a shared input directory generated once with a fixed seed and used by both the Fortran and JAX executions. The JAX driver calls the generated bridge entry points. Inputs are supplied to the bridge in the Fortran layout, \texttt{(ncol, nz)}, and the bridge is responsible for converting them to the layout expected by the translated JAX kernels. The driver therefore exercises the complete translated execution path, including initialization, bridge-level layout conversion, JAX kernel execution, and conversion of outputs back to the Fortran layout. The driver writes the resulting variables, including \texttt{theta}, \texttt{qv}, \texttt{qc}, \texttt{qr}, \texttt{precl}, and \texttt{relhum}, to the driver output directory, along with diagnostic information and optional plots. This stage does not itself compare the JAX results against the Fortran reference. Instead, it produces the JAX output files that are consumed by the following numerical comparison stage.

\textbf{Numerical comparison against Fortran:}
The final validation stage compares the JAX driver outputs against the
Fortran reference outputs. For each output variable $v$, the comparison
stage reports the mean absolute error, the relative root-mean-square
error, and the maximum absolute difference $E_v$ (Eq.~\eqref{eq:maxdiff})
between the translated JAX implementation and the Fortran reference. The
comparison stage also applies physics-sanity checks to both the Fortran
and JAX outputs. These checks verify that potential temperature remains
within a physically plausible range and that mixing ratios,
precipitation, and relative humidity remain non-negative. Applying these
checks to the Fortran outputs as well as the JAX outputs helps confirm
that the reference data used for comparison is itself physically
reasonable. In addition, the comparison stage checks whether the
comparison itself is trustworthy by verifying that the Fortran and JAX
arrays have matching shapes.

\textbf{Acceptance criterion:}
A translation is accepted when all structural and physics checks pass and
it differs from the Fortran reference by no more than the Fortran
reference differs from itself when compiled in different ways. This follows the port-validation strategy of \cite{rosinski_accumulation_1997}, who validates a code ported to a new computing environment by showing that its departure from the original is no larger than the departure the original itself produces from
rounding-level initial perturbations. We apply the same principle to a
single call of the scheme, with recompilation rather than perturbation as
the source of rounding-level differences. For this we compute $E_v$, which corresponds to the maximum absolute difference between the translation and the Fortran reference defined in Equation~\ref{eq:maxdiff}. 
To measure the variation of the reference itself, we built the same
Fortran source, on the same inputs, in 22 configurations: GNU, Intel,
NVIDIA and Cray Fortran, at optimization levels from \texttt{-O0} to
\texttt{-O3} or the vendor's \texttt{-fast}, with and without fused
multiply-add, with strict and with each vendor's default floating-point
model, on CPU and, for the NVIDIA compiler, on an A100 GPU through
OpenACC. Every build reads bit-identical inputs and was
deterministic over repeated runs. The 22 builds produce 15 distinct sets of outputs. Writing
$x^{a}_{v,i}$ for the value of the variable $v$ at the grid point $i$ in the build
$a$, we define the \emph{envelope} of variable $v$, denoted $\varepsilon_v$, as the largest maximum absolute difference between any two builds,
\begin{equation}
  \varepsilon_v = \max_{(a,b)} \; \max_i \,\bigl| x^{a}_{v,i} - x^{b}_{v,i} \bigr| ,
  \label{eq:envelop}
\end{equation}
where the outer maximum runs over all 231 unordered pairs $(a,b)$ of the
22 builds. A translation is accepted when all structural and physics
checks pass and, for every output variable $v$,
\begin{equation}
  E_v \le \varepsilon_v .
  \label{eq:accept}
\end{equation}

The envelope is not computed by the framework. It is obtained
once per scheme by a separate procedure that compiles the Fortran
reference under a set of compiler configurations, runs each build on the
same inputs, and compares the outputs pairwise; the resulting per-variable
values are supplied to the comparison stage as its thresholds. 
The envelope is specific to one input set, one call of the scheme, and one machine. The compilation ensemble characterizes numerical variability for the tested inputs, it does not establish coverage of branches triggered by special values, which require additional targeted tests. In a time-integrated run, these round-off differences would grow, for the Fortran builds as well as for the translation. Preliminary work is being done to establish how well LAFT translated code integrates into existing atmospheric frameworks such as EAMxx \cite{e3sm_project_energy_2024,donahue_exascale_2024}; initial phases of this work include comparing computational performance and accuracy between the original Fortran and the LAFT translated JAX implementation. 

\subsubsection{LLM Repair}
\label{sec:llmrepair}
Each validation stage produces its own result artifact, which is overwritten on every run so that it reflects the latest validation attempt. The linter records a per-check pass/fail list, diagnostic messages, and an aggregate score. Runtime validation records module import status, wrapper and core execution status, finite-output checks, and the full Python traceback for any exception. The driver and comparison stages emit structured results from the end-to-end execution and numerical comparison. In particular, the comparison result records shape agreement, output freshness, physics-sanity checks on both the Fortran reference and the JAX translation, and per-variable MAE values. This allows failures to be attributed either to a genuine numerical disagreement or to a problem in the comparison setup itself. Above these per-stage artifacts, LAFT maintains a structural file, \texttt{\detokenize{workflow.state.json}}, that records the repair log history. When a validation stage fails, the workflow state advances to the corresponding failed state and appends an entry to the repair log. Each repair-log entry records the failure stage, an error summary, the edited file, and the change applied. This log drives a bounded repair loop: each repair is applied directly to the final translated file. After each repair, LAFT reruns the validation ladder from linting forward. The loop is capped at five attempts. If validation still fails, the framework marks the run as aborted and writes a durable report containing the complete repair history and the final unresolved error. Thus, even unsuccessful translations remain auditable. For Qwen2.5-32B, however, the initial four-pass translation is generated by Qwen through its API, while Claude Code orchestrates validation and applies any repairs.

\subsubsection{Profiling and Feedback Loop}
\label{sec:profiling}
After numerical validation, LAFT evaluates whether the translated JAX code uses the GPU efficiently. Each optimization iteration consists of four phases: profiling, diagnosis, repair, and revalidation. LAFT records profiling history and diagnostic outcomes in a persistent state file, \texttt{\detokenize{profile_state.json}}, so that performance analysis remains auditable across iterations. The profiler traces the array-intensive routine using two problem sizes, 1,000 and 250 columns, each with 56 vertical levels. Comparing these two sizes helps distinguish vectorized execution from implementations whose runtime scales sequentially with the number of columns. LAFT also inspects the optimized XLA representation using a lightweight parser that counts fusion, scatter, gather, and \texttt{\detokenize{while}} instructions. 
Runtime traces captured with \texttt{jax.profiler.trace} measure device
execution time, end-to-end execution time, dispatch overhead,
device-to-host copy events, and the number of times each XLA
\texttt{while} instruction is entered during a call. These loop-entry
counts help distinguish batched column execution from sequential
column iteration: the batched implementations enter the subcycle loop
once per call, whereas the retained sequential Qwen implementation
enters it once per column. 
A local diagnostic engine converts the trace and HLO measurements into prioritized, machine-readable findings. Three defect classes determine production readiness. First, a sequential column loop is classified as critical, because it prevents the computation from exposing column-level parallelism to XLA. The prescribed repair replaces column-wise iteration with a batched formulation, such as \texttt{\detokenize{jax.vmap}}. Second, compiled HLO scatter instructions are classified as high severity because irregular writes can inhibit fusion. The prescribed repair replaces indexed updates with full-array masked operations, such as \texttt{\detokenize{jnp.where}}. Gather operations are treated differently from scatter operations. A scatter represents an indexed write, which can inhibit fusion and introduce irregular memory updates, so compiled scatter instructions are treated as high-severity performance defects. A gather, by contrast, represents an indexed read. Indexed reads may be necessary in scientific kernels, for example when a value at one vertical level depends on neighboring levels above or below it. Therefore, gathers that are absorbed into fused XLA regions are accepted. Unfused gathers are reported only as non-gating advisories: they may indicate an avoidable performance issue, but they do not by themselves make the translation unacceptable. Additional measurements, including instructions per fusion and execution time, are reported for characterization but do not by themselves determine the readiness verdict. After every profiling-driven repair, LAFT reruns the complete numerical-validation chain before profiling continues. This prevents performance-oriented changes from silently breaking scientific correctness. The feedback loop terminates when no actionable profiling issues remain or when the repair budget is exhausted. The repair budget is exhausted when the same issue persists through five consecutive repair attempts. If the issue concerns numerical correctness, the run is marked as aborted. If it concerns GPU performance, an implementation that passes numerical validation may be retained for experimental analysis, with the unresolved performance limitation explicitly reported.
\section{Case Study and Experimental Setup }
\subsection{Kessler Microphysics}
The source Fortran code used in the experiments implements the Kessler warm-rain microphysics scheme from the ESCOMP atmospheric physics repository \cite{escomp_ccpp-enabled_2024}. Kessler warm-rain microphysics is a classical bulk microphysics parameterization introduced by Kessler \cite{kessler_distribution_1969}. It represents warm-cloud processes using three water species: water vapor, cloud water, and rainwater, with no ice categories. In this implementation, the scheme is called once per vertical column and updates temperature and moisture after each model time step. Kessler provides a compact but representative benchmark for LAFT because it combines scientific, translation, and performance challenges in a form that can be validated in detail. The scheme exposes natural parallelism across atmospheric columns, but also contains vertical loops, threshold-based conditional updates, time-subcycling logic, array updates, and multiple interacting prognostic and diagnostic variables. These features test whether an LLM-generated translation can preserve the Fortran reference behavior while expressing the computation in a form suitable for JAX/XLA optimization. Kessler also exercises the bridge layer because the translated kernel must preserve the Fortran-facing interface while operating under the JAX layout expected by the generated GPU code. Thus, although Kessler is smaller than many production parameterizations, it captures key challenges that appear in larger atmospheric physics schemes: layout conversion, loop restructuring, control-flow translation, numerical validation, and GPU performance sensitivity.

\subsection{Implementations Evaluated}
\label{sec:implementations}

Four independently generated JAX translations were evaluated, produced by the
following LLMs: Claude Sonnet 4.6, GPT 5.4 Thinking, Gemini 3.1 Pro, and
Qwen2.5-32B. All models received the same four-pass translation prompts and
produced a JIT-compiled computational core with a Python wrapper. The three
proprietary models were accessed through their vendor interfaces with default
settings. Qwen2.5-32B was served locally via vLLM with a temperature of 0.1,
all remaining sampling parameters (\texttt{top\_p}, \texttt{top\_k},
\texttt{min\_p}, \texttt{repetition\_penalty}) at vLLM defaults. Thinking mode was disabled, no RoPE scaling was used, and an adaptive output cap was applied within a context length
of \texttt{max\_model\_len} = 22{,}528 tokens.

The JAX implementations were compared with the original serial Fortran in
terms of performance and accuracy. An OpenACC version was also included in the
scalability analysis as a manually accelerated GPU reference. All calculations
used 64-bit floating-point arithmetic. For the scalability experiments,
physically structured inputs were generated using a fixed random seed. The number of atmospheric columns ranged from 50 to $10^{6}$,
while the number of vertical levels remained fixed at 56.
\subsection{Hardware/Software Platform}
The GPU experiments were conducted on the NCAR Derecho supercomputer using a single NVIDIA A100-SXM4 GPU with 40 GB of memory. The recorded system used CUDA 13.0, Python 3.10.19, JAX 0.6.2.
Fortran source files were parsed using tree-sitter 0.20.4 with the Fortran grammar provided by tree-sitter-languages 1.10.2.
The serial Fortran reference used one CPU core and was compiled with gfortran -O2.
The OpenACC reference was compiled with nvfortran (NVIDIA HPC SDK 26.1) using -O3 -acc -mp=gpu.
\subsection{Metrics}
Numerical accuracy was evaluated for potential temperature, water
vapor, cloud water, rainwater, precipitation rate, and relative humidity.
In the final validation stage, the JAX driver outputs were compared
with the Fortran reference outputs. For each output variable $v$, the comparison stage reports $E_v$, the maximum absolute difference between the translation and the Fortran reference, defined in Eq.~\eqref{eq:maxdiff}:
\begin{equation}
  E_v = \max_i \,\bigl| x^{J}_{v,i} - x^{F}_{v,i} \bigr| ,
  \label{eq:maxdiff}
\end{equation}
where $x^{J}_{v,i}$ and $x^{F}_{v,i}$ are the JAX and Fortran values at
grid point $i$. The index $i$ runs over the 128~columns
$\times$ 56~levels of the test case for the three-dimensional fields and
over the 128~columns for the precipitation rate. The maximum absolute
difference $E_v$ is the quantity used by the acceptance criterion
(Section~\ref{sec:validation}). Additional physical checks verified that potential temperature remained within prescribed bounds and that moisture, precipitation, and relative humidity remained nonnegative. 

Performance was measured as the median wall-clock time of five calls
following one JIT warm-up call. Each timed call included the Python
invocation of the JIT-compiled function and a subsequent
\texttt{jax.block\_until\_ready()} call on the first returned output,
potential temperature, before the host timer was read. The reported
cached execution times therefore include dispatch overhead, device
execution, and synchronization, while excluding JIT compilation, the
initial host-to-device input transfers, and device-to-host output
transfers. For each model and column count, one input state generated
with seed 42 was reused for the warm-up and all five timed calls.

Data-layout conversion and host--device transfer costs were measured separately using the bridge benchmark. Speedup was calculated relative to the serial Fortran execution time. 

Profiling was performed in separate runs from the scalability timing experiments. Standard deviations across the five timed calls were calculated. Scalability was evaluated for column counts ranging from $50$ to $10^{6}$.
\section{Experiment 1: Execution Design and Comparison of LLM Translation Performance}
This experiment evaluates whether independently generated LLM translations of the Kessler microphysics routine produce executable, numerically faithful, and GPU-suitable JAX implementations. We compare four translated implementations produced by Claude, GPT, Gemini, and Qwen, as described in Section~\ref{sec:implementations}, against the original Fortran reference. The analysis considers both scientific correctness and execution design. All reported results correspond to numerically validated implementations
after the performance-repair stage, where applicable. For Qwen, the sequential column loop remained unresolved after five consecutive repair attempts, and the retained implementation is included to characterize its execution design and GPU performance.

\subsection{Execution Design and XLA Trace Behavior}
Although all four translations passed the static JAX-quality checks, their generated execution designs differed. Tables~\ref{tab:primitive-choices} and~\ref{tab:jax-trace-summary}
provide complementary views of the translated implementations. Table~\ref{tab:primitive-choices} summarizes the code design choices,
including column loop, level updates, top boundary handling,
and subcycle loop. 
Table~\ref{tab:jax-trace-summary} reports counts for one traced call to the JIT-compiled \texttt{kessler\_run\_core} function per model, following one untraced warm-up call. Inputs are already device-resident, and the trace interval ends after the outputs are ready, excluding compilation and initial input transfers. 
The ``XLA operations'' column counts executed XLA-level operations
on the device stream, including fusions, calls, and transposes,
but excluding memory copies. The ``Unique operations'' column
counts distinct operation identities in the trace. The
``While-loop executions'' column reports the maximum number of
entries into any individual XLA \texttt{while} instruction during
the call.
Together, the tables connect the structure of the generated code with its compiled execution behavior at $ncol=1000$, $nz=56$, and $dt=60$ s. The results demonstrate that GPU performance is not determined only by whether a translation uses JAX, but by how the generated code exposes parallelism and array operations to XLA. The Qwen-generated implementation reported in the tables and figures passed numerical validation but retained a sequential column loop. The sequential column loop remained unresolved after five consecutive performance-repair attempts targeting this issue. We therefore report the retained, numerically validated version to document its execution design and performance limitation.
\begin{table}[htbp]
\centering
\caption{Primitive choices used by the translated implementations. Claude and GPT are grouped because they produced the same design choices for these aspects.}
\label{tab:primitive-choices}
\begin{tabular}{llll}
\toprule
\textbf{Design aspect} & \textbf{Claude/GPT} & \textbf{Gemini} & \textbf{Qwen} \\
\midrule

Column loop &
\texttt{vmap} &
\texttt{vmap} &
\texttt{fori\_loop} \\

\addlinespace[0.6em]

Level loop &
\begin{tabular}[c]{@{}l@{}}
\texttt{jnp.arange}
\end{tabular} &
\begin{tabular}[c]{@{}l@{}}
\texttt{jnp.arange(nz) + clip}
\end{tabular} &
\begin{tabular}[c]{@{}l@{}}
\makecell[l]{\texttt{jnp.arange}\\
\texttt{.at[klevs].set}}
\end{tabular} \\

\addlinespace[0.6em]

Top boundary &
scalar index &
\begin{tabular}[c]{@{}l@{}}
\texttt{jnp.clip + jnp.where}
\end{tabular} &
\begin{tabular}[c]{@{}l@{}}
\texttt{.at[].set()}
\end{tabular} \\

\addlinespace[0.6em]

Subcycle loop &
\texttt{while\_loop} &
\texttt{while\_loop} &
\texttt{while\_loop} \\
\bottomrule
\end{tabular}
\end{table}

\begin{table}[htbp]
\centering
\caption{Trace counts per warmed-up call to the translated Kessler core
at $ncol=1000$, $nz=56$, and $dt=60$ s.}
\label{tab:jax-trace-summary}
\begin{tabular}{lrrr}
\toprule
\textbf{Model}
& \makecell{\textbf{XLA}\\\textbf{operations}}
& \makecell{\textbf{Unique}\\\textbf{operations}}
& \makecell{\textbf{While-loop}\\\textbf{executions}} \\
\midrule
Claude & 292       & 39 & 1              \\
GPT    & 292       & 39 & 1              \\
Gemini & 138       & 25 & 1              \\
Qwen   & 190{,}546 & 38 & 1{,}000        \\
\bottomrule
\end{tabular}
\end{table}
Among the three column-vectorized translations, the main difference is how each model expresses vertical-level operations (Table~\ref{tab:primitive-choices}). Claude and GPT use \texttt{jax.vmap} across columns and subset-based \texttt{jnp.arange} expressions over selected vertical levels. Their vertical updates require separate boundary-indexing logic and use gather--compute--scatter patterns. The traces reflect this execution structure: both implementations execute 292 XLA operations, with 39 unique operations (Table~\ref{tab:jax-trace-summary}).
Gemini also uses \texttt{jax.vmap} across columns, but applies operations over the full vertical array using \texttt{jnp.arange(nz)} and handles boundaries with \texttt{jnp.clip} and \texttt{jnp.where}. This avoids separate scalar boundary updates and reduces the need for scatter-style indexed updates. Its trace records 138 XLA operations and 25 unique operations, roughly half the XLA-operation count of Claude and GPT. These counts are consistent with a more compact execution structure and the lower cached execution times shown in Figure~\ref{fig:llmxfortran}.
In contrast, the Qwen implementation uses a sequential
\texttt{lax.fori\_loop} over columns, while vectorizing vertical-level
operations within each column using \texttt{jnp.arange} and indexed
\texttt{.at[klevs].set} updates. The subcycle loop is nested inside
the column loop and is therefore entered separately for each column,
producing 1,000 executions at $ncol=1000$.
The sequential column execution limits column-level parallelism
and is consistent with the substantially longer cached execution
times observed for Qwen.
\subsection{GPU Scaling of LLM-Generated JAX Translations}
\begin{figure}[htbp]
    \centering
    \includegraphics[width=0.85\linewidth]{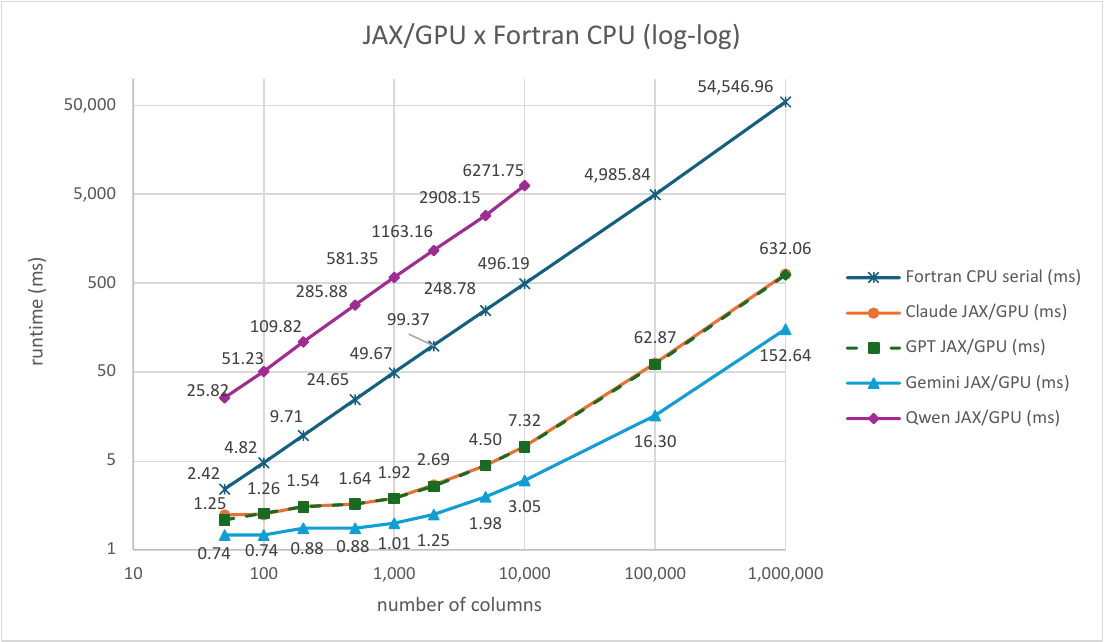}
    \caption{Cached runtime scaling of the serial Fortran baseline and LLM-generated JAX/GPU translations as a function of column count. The plot uses logarithmic axes. Claude, GPT, and Gemini scale substantially better than serial Fortran, with Gemini achieving the best performance among the numerically validated translations. The GPU scaling of the Qwen-generated translation is limited by sequential column execution.}
    \label{fig:llmxfortran}
\end{figure}

Figure~\ref{fig:llmxfortran} shows cached runtime scaling as the number of columns increases from 50 to ($10^6$), with the vertical dimension fixed. The serial Fortran baseline scales approximately linearly with column count, increasing from 2.4 ms at (ncol=50) to 54,546 ms at (ncol=$10^6$). This trend is expected for a serial CPU implementation, where additional columns are processed sequentially.
The JAX/GPU translations generated by Claude, GPT, and Gemini scale much more favorably. At small problem sizes, all three JAX implementations have a nearly flat runtime because GPU dispatch and fixed overheads dominate the cost. As $ncol$ increases, runtime grows, but at a much slower rate than serial Fortran. At $ncol=10{,}000$, Claude and GPT require 7.32 ms and 7.26 ms, respectively, compared with 496.194 ms for serial Fortran. 

Claude and GPT show nearly identical scaling behavior across the tested column counts; for readability, Figure~\ref{fig:llmxfortran} shows data labels only for the Claude curve in this overlapping pair, while both curves remain plotted. This corresponds to approximately $68\times$ speedup over the serial CPU baseline. Gemini is faster, requiring only 3.05 ms at $ncol=10{,}000$, or approximately $163\times$ speedup over serial Fortran.

At larger problem sizes, Gemini's advantage becomes more pronounced. At $ncol=100{,}000$, Gemini requires 16.30 ms, compared with 62.87 ms for Claude, 60.96 ms for GPT, and 4,985.835 ms for serial Fortran. At $ncol=1{,}000{,}000$, Gemini requires 152.64 ms, while Claude and GPT require 632.06 ms and 619.74 ms, respectively, compared with 54,546.96 ms for serial Fortran. At this largest problem size, the cached execution speedups over serial Fortran are approximately $357\times$ for Gemini, $86\times$ for Claude, and $88\times$ for GPT. Gemini is also approximately $4\times$ faster than the Claude and GPT translations. 
This behavior is consistent with the primitive-level analysis in Table~\ref{tab:primitive-choices}, where Gemini's full-array formulation with masking exposes more regular array operations to XLA, enabling better fusion and scaling.
Qwen shows a qualitatively different trend. Its runtime is already much slower than serial Fortran at small problem sizes and grows steeply with column count, reaching 6,271.75 ms at (ncol=10{,}000). This poor scaling is consistent with its use of \texttt{lax.fori\_loop} over columns, which preserves a more sequential execution structure and fails to expose the dominant column-level parallelism.
\subsection{Numerical Accuracy}
The numerical accuracy was evaluated by comparing the outputs of the validated JAX translations with the Fortran reference for the six output variables.

Table~\ref{tab:envelope} reports the observed Fortran-vs-Fortran
envelope and the worst JAX-vs-Fortran difference over the four
translations. The second column reports the largest maximum absolute
difference observed between any two tested Fortran builds. The third
column reports the largest maximum absolute difference between the
four JAX translations and the Fortran reference. A ratio below one
in the fourth column indicates that the translation error is smaller
than the observed Fortran-vs-Fortran envelope for that variable.

\begin{table} [H]
\centering
\caption{Fortran-vs-Fortran envelope $\varepsilon_v$ (largest maximum absolute difference
between any two builds of the reference) and worst JAX-vs-Fortran difference $E_v$ over the
four translations. Units: K for potential temperature, kg\,kg$^{-1}$ for the water vapor,
cloud water and rain water mixing ratios (with respect to dry air), m\,s$^{-1}$ of liquid
water for the precipitation rate, and \% for relative humidity.}
\label{tab:envelope}
\begin{tabular}{lccc}
\hline
Variable & $\varepsilon_v$ & $E_v$ & $E_v/\varepsilon_v$ \\
\hline
Potential temperature & $1.1\times10^{-13}$ & $5.7\times10^{-14}$ & 0.52 \\
Water vapor           & $7.1\times10^{-17}$ & $6.2\times10^{-17}$ & 0.87 \\
Cloud water           & $5.7\times10^{-17}$ & $2.9\times10^{-17}$ & 0.51 \\
Rain water            & $3.3\times10^{-17}$ & $2.1\times10^{-17}$ & 0.64 \\
Precipitation rate    & $1.4\times10^{-19}$ & $1.1\times10^{-19}$ & 0.79 \\
Relative humidity     & $9.8\times10^{-13}$ & $4.1\times10^{-13}$ & 0.42 \\
\hline
\end{tabular}
\end{table}
For every output variable, the maximum absolute difference between the
JAX translations and the Fortran reference lies within the observed
Fortran-vs-Fortran envelope. The ratios range from 0.42 to 0.87,
indicating that the translation differences are smaller than the
largest differences observed among the tested Fortran builds.
\subsection{Bridge Overhead}
Table~\ref{tab:bridge_overhead} reports the bridge-phase breakdown as a function of column count. LAFT measures the cost of moving data between the CPU and GPU separately from the cost of the main JAX kernel in order to isolate interface overhead from numerical computation. The measured phases are the host-to-device transfer (H2D), the input permutation performed inside the JIT-compiled computation, the corresponding output permutation, and the batched device-to-host transfer (D2H) that returns the results to the CPU.
\begin{table}[htbp]
\centering
\caption{Bridge phase breakdown by column count (\texttt{ncol}), measured
using the C-contiguous caller layout of the production driver. The two
permutation phases execute inside the JIT-compiled computation and are fused
into the production kernel. The phase sum (\emph{Total}) estimates bridge overhead from separately measured components; the overhead in the production call may differ
because XLA can optimize the permutations together with the computation. The final column reports the total for a column-major (Fortran-ordered) caller, for which the H2D phase also incurs a
host-side relayout copy.}
\label{tab:bridge_overhead}
\setlength{\tabcolsep}{4pt}
\renewcommand{\arraystretch}{1.12}
\begin{tabular}{@{}rrrrrrrrr@{}}
\toprule
\makecell{Columns}
& \makecell{Input\\(MB)}
& \makecell{Output\\(MB)}
& \makecell{H2D\\(ms)}
& \makecell{Perm. in\\(ms)}
& \makecell{Perm. out\\(ms)}
& \makecell{D2H\\(ms)}
& \makecell{Total\\(ms)}
& \makecell{F-order\\total (ms)} \\
\midrule
50            & 0.2    & 0.1    & 2.087   & 0.179 & 0.150 & 0.467   & 2.882    & 2.840    \\
100           & 0.4    & 0.2    & 2.041   & 0.197 & 0.149 & 0.504   & 2.890    & 3.064    \\
200           & 0.9    & 0.4    & 2.181   & 0.187 & 0.153 & 0.513   & 3.038    & 2.806    \\
500           & 2.2    & 1.1    & 2.220   & 0.192 & 0.160 & 0.566   & 3.141    & 2.934    \\
1{,}000       & 4.5    & 2.2    & 2.277   & 0.200 & 0.159 & 0.603   & 3.245    & 3.026    \\
2{,}000       & 9.0    & 4.5    & 2.307   & 0.186 & 0.136 & 0.857   & 3.498    & 3.487    \\
5{,}000       & 22.4   & 11.2   & 2.558   & 0.210 & 0.146 & 1.337   & 4.301    & 4.994    \\
10{,}000      & 44.9   & 22.5   & 4.387   & 0.243 & 0.165 & 2.364   & 7.129    & 11.656   \\
100{,}000     & 448.8  & 224.8  & 31.233  & 0.793 & 0.448 & 82.471  & 114.988  & 159.331  \\
1{,}000{,}000 & 4488.0 & 2248.0 & 319.061 & 6.813 & 3.728 & 706.927 & 1036.640 & 1468.771 \\
\bottomrule
\end{tabular}
\end{table}
The two permutation phases contribute very little to the total overhead. Even at $ncol=10^6$, the input and output permutations require only 6.8~ms and 3.7~ms, respectively. For problem sizes up to $10^4$ columns, the total bridge cost remains between approximately 2.9 and 7.1~ms. At larger sizes, the overhead grows with data volume, reaching 1,036~ms at $ncol=10^6$. Most of this cost comes from data movement: at $ncol=10^6$, the H2D phase requires 319~ms and the batched D2H phase requires 707~ms.

The table reports these measurements for the C-contiguous caller layout used by the production driver. When the caller instead supplies arrays in Fortran order, an additional host-side relayout is required before transfer, increasing the total to 1,468~ms at $ncol=10^6$. These results show that LAFT's layout-conversion overhead is small relative to the cost of host--device data movement. 
Because the phases are measured separately, their sum is an estimate
of bridge overhead rather than a direct measurement of the overhead
in the production call.
\section{Experiment 2: JAX/XLA versus OpenACC GPU}
\begin{figure}[htbp]
    \centering
    \includegraphics[width=0.85\linewidth]{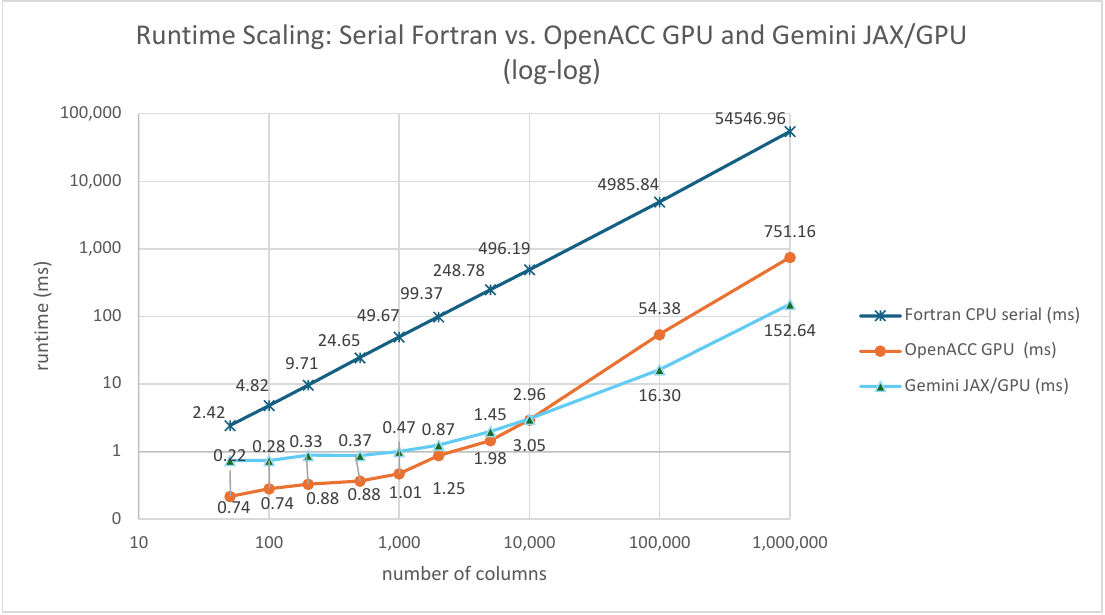}
    \caption{Runtime scaling of the serial Fortran CPU baseline, OpenACC GPU implementation, and Gemini-generated JAX/GPU translation as a function of column count. The plot uses logarithmic axes. OpenACC is faster at small and moderate problem sizes; the two GPU implementations are nearly equal at $ncol=10{,}000$; and Gemini JAX/GPU becomes faster at larger column counts.}
    \label{fig:fortran_openACC_gemini}
\end{figure}

Figure~\ref{fig:fortran_openACC_gemini} compares runtime scaling for three implementations of the Kessler kernel: the original serial Fortran CPU baseline, the manually accelerated OpenACC GPU implementation, and the Gemini-generated JAX/GPU translation. The plot uses logarithmic axes and varies the number of columns from 50 to ($10^6$), while holding the vertical dimension fixed. This comparison distinguishes two performance questions: how much GPU acceleration improves over serial Fortran, and how the LLM-generated JAX/XLA implementation compares with a manually accelerated OpenACC GPU version.
Both GPU implementations substantially outperform the serial Fortran baseline across the tested range. The serial Fortran runtime increases nearly linearly with column count, from 2.42 ms at ($ncol=50$) to 54,546.96 ms at ($ncol=10^6$). In contrast, both OpenACC and Gemini JAX remain below 1 ms for the smallest cases and scale much more favorably as the column count grows.

At small and moderate problem sizes, OpenACC is faster than Gemini JAX in kernel-only execution. For example, at ($ncol=50$), OpenACC requires 0.218 ms, compared with 0.74 ms for Gemini JAX. At ($ncol=10{,}000$), the two GPU implementations become nearly equal, with OpenACC requiring 2.96 ms and Gemini requiring 3.05 ms. At larger problem sizes, the scaling trend reverses: Gemini JAX becomes faster than OpenACC, requiring 16.30 ms versus 54.38 ms at (ncol=100{,}000), and 152.64 ms versus 751.16 ms at ($ncol=10^6$).
This crossover suggests that OpenACC has lower overhead and better performance for smaller column counts, while the Gemini-generated JAX/XLA implementation benefits more strongly from large-batch execution. Once the problem size is large enough to amortize JAX/XLA overheads, Gemini’s fused array formulation scales better than the OpenACC GPU implementation. Thus, the result does not imply that one GPU programming model is uniformly superior; instead, it shows that the performance advantage depends on problem size and on how effectively each implementation exposes parallelism and fusion opportunities.

\section{Future Work}
A future direction is to use profiling results not only for post-translation repair, but also to improve the prompt policy itself. 
The results associate Gemini's full-array masked formulation with a
more compact XLA execution trace and better GPU scaling than the
subset-based gather--compute--scatter formulations and the sequential
column implementation evaluated here.
Future LAFT versions could feed this design rule back into the translation prompts, explicitly encouraging full-array masking and discouraging scatter-heavy updates or sequential column iteration.
Another direction is to extend LAFT to larger and more complex atmospheric physics parameterizations. Ongoing experiments with the P3 microphysics scheme and the Droplet scheme are being used to test how the workflow scales beyond Kessler. These cases introduce additional challenges, including more procedures, richer control flow, stronger cross-procedure dependencies, lookup tables, sedimentation logic, and more complex bridge and validation requirements.

Ongoing and future work also includes integrating the LAFT generated bridge and physics implementations into existing atmospheric modeling frameworks. Preliminary studies indicate that the LAFT translated JAX implementation of Kessler bridged into an EAMxx atmospheric process has comparable complexity and performance as the original Fortran Kessler implementation.

\section{Conclusion}
This paper presented LAFT, an LLM-assisted framework for translating legacy Fortran atmospheric physics code into GPU-accelerated JAX. LAFT addresses the translation problem as more than a prompt-to-code task: it combines deterministic Fortran analysis, bridge-based memory-layout conversion, multi-pass LLM translation, automated validation and repair, and profiling-guided performance analysis. Using the Kessler warm-rain microphysics scheme as a case study, we showed that current LLMs can generate JAX translations that reproduce the Fortran reference to near floating-point precision for the evaluated variables. Claude, GPT, Gemini, and Qwen achieved MAE values in the range of approximately $10^{-20}$--$10^{-15}$, indicating that LLM-assisted translation can preserve the numerical behavior of this atmospheric physics kernel when embedded in a validation-driven workflow. At the same time, the experiments show that numerical correctness alone is not sufficient. The validated translations differed substantially in GPU performance because the models produced different execution designs. 
Claude and GPT produced correct translations that exposed column-level
parallelism, but their subset-based vertical updates introduced
gather--compute--scatter patterns and a larger number of executed XLA
operations. Gemini produced a full-array formulation using clipping and
masking, which yielded a more compact XLA execution trace and achieved
the best scaling. 
Qwen also achieved numerical agreement with the Fortran reference, but preserved a sequential column loop and failed to expose the dominant source of parallelism, leading to poor GPU scaling. The comparison with OpenACC further illustrates the importance of problem size and execution structure. OpenACC performs better for small and moderate column counts, while the Gemini-generated JAX/XLA implementation becomes faster at larger column counts, where large-batch execution and fusion benefits dominate. This crossover highlights that JAX/XLA can be competitive with manually accelerated GPU code, but only when the generated program exposes the right structure to the compiler. The results suggest that future LLM-assisted scientific-code translation systems should treat correctness validation and performance profiling as first-class components of the workflow.
%%%%%%%%%%%%%%%%%%%%%%%%%%%%%%%%%%%%%%%%%%%%%%%
%
% DATA SECTION and ACKNOWLEDGMENTS
%
%%%%%%%%%%%%%%%%%%%%%%%%%%%%%%%%%%%%%%%%%%%%%%%

\section*{Open Research Section}
The LAFT framework source code and reproduction artifacts used in this study are archived on Zenodo at https://doi.org/10.5281/zenodo.22696825
\section*{Conflict of Interest declaration}
The authors declare there are no conflicts of interest for this manuscript.
\section*{Acknowledgments}
This material is based upon work supported by the NSF National Center for Atmospheric Research, which is a major facility sponsored by the U.S. National Science Foundation under Cooperative Agreement No. 1852977. We would like to acknowledge high-performance computing support from Derecho and Casper \cite{noauthor_ncar_2026, noauthor_ncar_2026-1} provided by the Computational and Information Systems Laboratory, NCAR, and sponsored by the National Science Foundation. 

\bibliography{references}

\end{document}